\documentclass[12pt]{article}
\usepackage{epsf}

\newcommand{\bee}{\begin{equation}}
\newcommand{\ee}{\end{equation}}
\newcommand{\bea}{\begin{eqnarray}}
\newcommand{\eea}{\end{eqnarray}}
\newcommand{\R}{\rm I\kern-.2emR}
\newcommand{\C}{\rm \kern.25em\vrule height1.4ex
depth-.12ex width.06em\kern-.31em C}
\newcommand{\N}{{\rm I\kern-.16em N}}
\newcommand{\Z}{{\rm Z\kern-.35em Z}}

\newcommand{\NP}{{\sl Nucl. Phys.}}
\newcommand{\PL}{{\sl Phys. Lett.}}
\newcommand{\CMP}{{\sl Commun. Math. Phys.}}
\newcommand{\PRL}{{\sl Phys. Rev. Lett.}}
\newcommand{\PR}{{\sl Phys. Rev.}}
\newcommand{\AP}{{\sl Ann. Phys. (N.Y.)}}
\begin{document}
  \begin{flushright}
MPI-Ph/91-88 \\
October 1991
(slightly revised)
\end{flushright}
\bigskip\bigskip\begin{center}
{\bf
The Difference between Abelian and Non-Abelian Models:\\ Fact
and Fancy
}\footnote {This  paper reflects the state of affairs as it was in 1991. 
In some areas there have been important developments since then.}
\end{center}
\centerline{Adrian Patrascioiu}
\centerline{\it Physics Department, University of Arizona,}
\centerline{\it Tucson, AZ 85721, U.S.A.}
\centerline{\it e-mail: patrasci@ccit.arizona.edu}
\vskip5mm
\centerline{and}
\centerline{Erhard Seiler}
\centerline{\it Max-Planck-Institut f\"ur Physik}
\centerline{\it (Werner-Heisenberg-Institut)}
\centerline{\it F\"ohringer Ring 6, 80805 Munich, Germany}
\centerline{\it e-mail:ehs@mppmu.mpg.de}
\bigskip \nopagebreak

\begin{abstract}
The commonly accepted belief that non-Abelian and Abelian models are
different because of the presence/absence of instantons and/or perturbative     
asymptotic freedom is analyzed from a historical perspective. The presentation  
covers the major developments which brought about this dogma, as well as all    
the supportive evidence produced since. For a model possessing both asymptotic  
freedom and instantons it is shown rigorously that a disorder variable varies   
nonanalytically with the temperature.
\end{abstract}
\newpage
\centerline{{\bf 1.Introduction}}                                               
\vskip5mm
The generally accepted belief among condensed matter and particle               
physicists is that there is a dramatic difference between Abelian nonlinear     
$\sigma$-models in two dimensions ($2D$) and gauge theories in $4D$ and their   
non-Abelian counterparts. This opinion, which did not exist prior to            
roughly 1973, has its origin in two important observations:                     
                                                                                
i) the importance of topological properties                                     
(Kosterlitz and Thouless \cite{KT})                                           
                                                                                
ii) the discovery of perturbative asymptotic freedom (Gross and Wilczek         
 \cite{GroWi} and Politzer  \cite{Poli}).                                   
                                                                                
The difference between the Abelian and the non-Abelian models is supposed to    
stem from the presence of instantons and/or asymptotic freedom in the latter,   
but not in the former. In the previous sentence we used `and/or' deliberately,  
to emphasize that one property does not entail the other, hence one may         
legitimately ask, which property, if any, is responsible for this accepted      
difference.                                                                     
                                                                                
We are not aware of any paper addressing this issue. Instead many papers        
report evidence in favor of what we shall call the orthodoxy, coming from all   
sorts of sources. For instance, it is argued that $QCD_4$ must be the correct   
theory of strong interactions because there is general agreement between its    
predictions and experiment (for a recent example see \cite{Nature}).          
In reality the confrontation of $QCD_4$ with                                    
experiment is nowhere near that of say $QED_4$, for two reasons:                
                                                                                
i) Any realistic computation must address the issue of constructing physical    
states (confinement, etc), which even $QCD$ believers admit is beyond their     
present technical abilities. Thus the best $QCD$ predictions involve, besides   
some well defined perturbative computation, some ad hoc phenomenological        
assumptions. Hence the situation is not at all like computing the electron $g-2$
factor in  $QED$.                                                               
                                                                                
ii) Even with all these uncertainties, the overall agreement                    
between                                                                         
theory and experiment is at the $10\%$ level,                                   
rather than $10^{-8}$, as in $QED_4$.                                           
Under these circumstances, to employ nature as an analog computer for learning  
the true properties of $QCD_4$ seems rather cavalier.                           
                                                                                
Of course nowadays the computer offers the opportunity to investigate           
numerically the properties of many models of interest. Numerous such studies    
have been performed and the general consensus is that they do corroborate       
theoretical expectations. Finally many people have tried to produce indirect    
evidence in favor of the orthodoxy by appealing to so-called exact solutions,   
the $1/N$ expansion, etc..                                                      
                                                                                
In this paper we offer a critical analysis of all the                           
evidence supporting the idea that Abelian                                       
and non-Abelian models have fundamentally different                             
properties. Without advocating the opposite point of view, we recall rigorously 
established facts and thus attempt to put all the supportive evidence in its    
proper perspective. After giving a historical overview of the subject           
we begin by recalling what is known about perturbation                          
theory and semiclassical (instanton) approximations. We then review the more    
recent findings regarded as supportive of the orthodoxy, including Monte        
Carlo studies. Our general conclusion is that the original belief that          
the presence of instantons and/or asymptotic freedom explains the difference    
between Abelian and non-Abelian models is unfounded. The same conclusion applies
to the more recent circumstantial evidence. Finally we argue that numerical     
studies are at best inconclusive -- one may even claim that they indicate       
the opposite. Also in this paper we give an outline of a rigorous proof         
that a certain modification of the nonlinear $O(N)$ models in $2D$,             
which retains asymptotic freedom and instantons,                                
has a nonanalytic change in its behaviour with the inverse temperature $\beta$  
(a similar modification and result were proved by Mack and Petkova for gauge    
theories \cite{MP}).                                                          
                                                                                
\vskip1cm                                                                       
\centerline{{\bf 2.Historical Background}}                                      
\vskip5mm
In 1966 Mermin and Wagner \cite{MW} introduced their celebrated theorem       
about the absence of symmetry breaking in $2D$ models enjoying invariance under 
a continuous, compact group. Initially it was thought that that result implied  
that $2D$ nonlinear $O(N)$ $\sigma$-models cannot undergo phase transitions,    
since the theorem says that they could not exhibit long range order (l.r.o.).   
Shortly thereafter, however, Stanley and Kaplan \cite{SK}                     
examined the high temperature expansion of the                                  
magnetic susceptibility in such models. Although one cannot rigorously compute  
the radius of convergence of a series by knowing only a finite number of terms, 
the numbers suggested very strongly that the magnetic susceptibility diverged   
at finite $\beta$. To reconcile such a property with the absence of l.r.o.,     
the term algebraic order was coined and initially it was believed that all      
$O(N)$ models with $N>1$ possess such a phase at sufficiently low temperatures. 
                                                                                
This paradigm survived until 1973, when the seminal paper of Kosterlitz         
and Thouless appeared \cite{KT}.                                              
They introduced the notion of topological order;                                
specifically they argued that at low temperature, the typical configurations    
of the system will be low lying excitations of the configuration of lowest      
energy. For $O(2)$ they will consist of                                         
bound vortex-antivortex pairs, since the energy of just one vortex diverges     
as $\ln L$ ($L$-linear size of the lattice); for $O(3)$ one can form instantons,
whose energy is $O(L^o)$. (We use the term instanton,                           
which was invented only later,                                                  
in 1976; the properties of the instanton configuration were described and       
employed by Kosterlitz and Thouless in Sect.6 of their 1973 paper.) Since the   
entropy of these topological defects is $O(\ln L)$ (the position of their       
center), in the $O(2)$ model, at low temperature, it will be overwhelmed by     
the energy and, Kosterlitz and Thouless argued, the model will enjoy            
topological order. On the contrary, in the $O(3)$ model the entropy will win    
and thus correlations will exhibit exponential decay even at low temperature.   
                                                                                
The proposal of Kosterlitz and Thouless pertained only to spin                  
systems and was not readily applicable to particle physics. However the same    
year particle physicists became very excited by their own discovery, namely     
that of asymptotic freedom by                                                   
Gross and Wilczek \cite{GroWi} and Politzer \cite{Poli}. The proof that in  
perturbation theory non-Abelian gauge theories could enjoy this property        
promised a resolution to the long sought field theoretic explanation of         
strong interactions and even the grand unification of weak, electromagnetic     
and strong interactions (GUTS). Moreover it was claimed that asymptotic         
freedom suggests naturally why quarks are confined: if the interactions are     
getting weaker at short distances, they must be getting stronger as one tries   
to separate a quark from an antiquark, hence asymptotically free theories       
should also be confining \cite{StW}.                                          
                                                                                
For approximately two years the developments in condensed matter and            
particle physics remained largely separated. However in 1975-1976 in papers     
destined to become classics, Polyakov \cite{Polyakov},                        
Belavin et al \cite{Belavin} and                                              
Br\'ezin and Zinn-Justin \cite{BZJ} bridged that gap.                         
Namely Polyakov and Br\'ezin and                                                
Zinn-Justin proved that the $O(N)$, $N\geq 3$, $2D$ nonlinear                   
$\sigma$-models are asymptotically                                              
free in perturbation theory, while Belavin, Polyakov, Schwartz and              
Tyupkin \cite{BPST} showed that                                               
Yang-Mills theories in $4D$ also possessed instantons -- just as the $O(3)$ spin
model in $2D$. Through these realizations a new paradigm was born: there exists 
a fundamental difference between Abelian and non-Abelian models, stemming from  
their different topological properties and/or existence or absence of           
asymptotic freedom.                                                             
                                                                                
This fifteen year old dogma remains unproven, yet it is widely believed.        
We will analyze its merits in the next section, but first we would like to      
finish our historical recollection. During the late 1970s several attempts      
were made to develop a semiclassical approximation based upon the instanton     
idea in both $2D$ $\sigma$-models and $4D$ gauge theories. In spite of the      
technical brilliance of these papers, Patrascioiu \cite{Adr} argued that      
the original computations by t'Hooft \cite{tH}                                
(for gauge theories) and Berg and L\"uscher \cite{BL}                         
and Fateev, Frolov and Schwartz \cite{FFS}                                    
(for $\sigma$-models) were incorrect and that                                   
uncontrollable infrared divergences plagued the semiclassical approximation.    
This fact was proven by Patrascioiu and Rouet for both gauge theories           
\cite{PatRouet} and $\sigma$-models \cite{PaRo}.                            
                                                                                
These mathematical facts, established in 1981, remain largely ignored by the    
community, which still talks about the resolution of the $U(1)$ problem by      
instanton effects or about the strong $CP$-violation problem caused by          
instantons. In reality                                                          
the $U(1)$ problem does not exist because the corresponding axial               
current has the famous Adler-Bardeen anomaly and nobody has ever shown that     
there exists some conserved, gauge invariant axial $U(1)$ current in the        
physical sector.                                                                
As for the `strong $CP$ problem', no such problem existed prior to 1976 and     
its origin is intimately connected with the belief that instanton effects       
force $QCD_4$ to have many $\theta$-vacua; since all but one such vacuum would  
be $CP$ violating, it was concluded that without 'fine tuning', the theory      
would not exhibit $CP$ invariance. The axion was introduced in 1977             
\cite{PQ} precisely to eliminate this need for fine tuning.                   
Extensive experimental searches have repeatedly failed to find this elusive     
particle. Is this an embarassement for $QCD_4$ ? Not in our opinion.            
The original instanton motivation was discredited by the infrared divergences   
found. Moreover the better understood $QED_2$, which also has topolgically      
nontrivial fields, revealed that while `$\theta$-vacua' can be defined by       
choosing suitable boundary conditions, there is no necessity to have $\theta$   
different from 0 \cite{CJS,FrSe}.                                            
                                                                                
A rigorous result supportive of the orthodoxy was obtained the same year.       
Fr\"ohlich and Spencer \cite{FS} proved rigorously                            
that Abelian models do undergo                                                  
the phase transition predicted by Kosterlitz and Thouless. As far as direct     
evidence goes, this can be considered the only positive result ever found and   
everything else seemed to cast doubt rather than support the orthodoxy.         
For instance after a real tour-de-force Bricmont, Fontaine, Lebowitz, Lieb and  
Spencer \cite{BFLLS} proved that in Abelian                                   
models ordinary perturbation theory does                                        
provide the correct asymptotic expansion in powers of $1/\beta$ for             
$\beta\to\infty$                                                                
(the proof is for an infinite lattice model, so it does not concern constructing
a continuum limit but only a thermodynamic limit). In spite of all their        
efforts (private communication to A.P. by J.Lebowitz), they could not extend the
proof to non-Abelian models. Of course the reader could consider this fact      
merely a measure of the prowess of mathematical physicists, however there are   
some exactly soluble cases, which we will discuss in the next section and they  
are not very reassuring -- perturbation theory produces (infrared) finite       
answers, which are correct in Abelian models and false (at $O(1/\beta^2$))      
in non-Abelian ones.                                                            
                                                                                
Another potential embarassment was pointed out by Solomon \cite{Sol}.         
By analyzing                                                                    
the high temperature expansion, he realized that if in the $O(N)$ model one     
changes the action from $S_i\cdot S_j$ to $(S_i\cdot S_j)^2$,                   
the series seems to predict a singularity at a                                  
finite positive $\beta$ for all $N<\infty$.                                     
This would be strange because this action leads also                            
to asymptotic freedom for $N\geq 3$. Two numerical studies triggered            
by Solomon's                                                                    
work \cite{FukugitaRP2},                                                      
\cite{Sinclair} observed a dramatic increase in                               
the correlation length and magnetic                                             
susceptibility at some finite $\beta$. Although these two studies produced      
practically identical data, in a classic example that beauty is in the eye of   
the beholder, Fukugita et al \cite{FukugitaRP2}                               
concluded that a transition to a massless phase                                 
was occurring while Sinclair \cite{Sinclair}                                  
claimed that it was just a cross-over regime.                                   
                                                                                
While one may easily dismiss numerical results, a rigorous result               
proving that asymptotic freedom could not be the source of the mass gap was     
obtained by Richard \cite{Richard}.                                           
Let us state it for simplicity for $O(3)$: consider                             
a modification of the measure such that $\sin\theta>\epsilon$                   
where $\theta$ is the                                                           
angle specifying the latitude. This modification destroys                       
$O(3)$ invariance and                                                           
instantons, but not asymptotic freedom. By using Ginibre's                      
inequalities Richard proved that provided $\epsilon>c/\sqrt\beta$ for some      
suitably chosen $c$, for $\beta$ sufficiently large the correlations of         
$S_x$ and $S_y$                                                                 
must decay algebraically. Inspired by this result, Patrascioiu  \cite{Pat}    
questioned the accepted relationship between the perturbative Callan-Symanzik   
$\beta$-function and the nature of the spectrum, pointing out several           
counterexamples.                                                                
                                                                                
As we stated in the introduction, the community has largely ignored all         
evidence shedding doubt upon the orthodoxy. Instead supportive evidence has     
been sought in many places. Since it is commonly argued that such supportive    
evidence has been found and that it is only a technicality that nobody has      
managed to prove rigorously that non-Abelian models are indeed fundamentally    
different from their Abelian counterparts, in the remainder of this paper we    
will analyze these findings in detail. We will also adapt the Mack-Petkova      
modification of gauge theories to $2D$ $O(N)$ nonlinear $\sigma$-models.        
For these models we will prove rigorously that although they are asymptotically 
free and possess instantons a certain disorder variable changes nonanalytically 
with $\beta$. Were it not for the fact that disorder variables are nonlocal     
observables in terms of the original spins, this would constitute a rigorous    
proof for the existence of a phase transition.                                  
\vskip1cm                                                                       
\centerline{{\bf 3.What is the Evidence for the Orthodoxy ?}}                   
\vskip5mm                                                                       
{\it a) Asymptotic freedom}                                                     
                                                                                
Asymptotic freedom was shown only                                               
in perturbation theory \cite{BZJ}, which                                      
is an expansion in small deviations                                             
from an ordered state. As stressed by                                           
Patrascioiu and Richard \cite{PatRich},                                       
the Mermin-Wagner theorem guarantees that                                       
such a state does not exist in $D\le 2$.                                        
Since in spite of this difficulty it was possible to prove                      
rigorously \cite{BFLLS} that in Abelian                                       
models perturbation theory does provide the                                     
correct asymptotic expansion, one could                                         
take the optimistic point of view                                               
that perturbation theory does work                                              
also in non-Abelian models and only                                             
technical difficulties are preventing a proof.                                  
However in $1D$, where one                                                      
knows the exact answer, one sees by explicit computation                        
that perturbation theory fails                                                  
for non-Abelian models, while giving correct                                    
answers for Abelian ones. (In fact since in $1D$ there                          
is a nonvanishing mass gap for any finite $\beta$, the infinite volume          
limit of the Green's functions is unique; yet in non-Abelian                    
models, perturbative answers depend upon the boundary                           
conditions used.)                                                               
                                                                                
In gauge theories we expect similar difficulties with perturbative              
predictions. Indeed Patrascioiu pointed out                                     
\cite{PatPRL54} that even after the gauge                                     
freedom has been completely eliminated,                                         
on an infinite lattice the fluctuations do not go to                            
zero as \hskip1cm                                                               
$\beta\to\infty$, as they do on a finite                                        
lattice.                                                                        
We would like to emphasize that this result, proven                             
in the complete axial gauge, is much                                            
stronger than Elitzur's theorem \cite{Elitzur}                                
on the impossibility to break spontaneously                                     
a gauge symmetry and in our opinion                                             
strongly suggests that perturbation theory                                      
is incorrect in non-Abelian gauge                                               
theories; this can be verified in $2D$                                          
(using periodic boundary conditions).                                           
\vskip1cm                                                                       
{\it b) Instanton computations}                                                 
                                                                                
We cannot delve here into the complicated instanton computations and refer the  
reader to the original papers quoted before (see also \cite{PAN}). We will    
only try to explain from a heuristic point of view why there is a problem. It   
is well known that a semiclassical approximation involves two steps:            
                                                                                
i) find a classical solution and                                                
                                                                                
ii) calculate Gaussian fluctuations around it                                   
                                                                                
\noindent                                                                       
The second step amounts to the calculation of the determinant of an operator    
which in the simplest case of one instanton takes the form \cite{tH}          
                                                                                
$${\cal O}=-\Delta^{-1}(-\Delta+V(x))=1-\Delta^{-1}V(x) \eqno(1)$$              
                                                                                
Here $V(x)$ is the operator of multipication by                                 
some instanton induced potential and the inverse of the                         
Laplacian appears to ensure proper normalization of                             
the functional integral.                                                        
The determinant has an ultraviolet divergence even in a finite                  
volume which can be cancelled by local counterterms (see for                    
instance \cite{Seiler,SeiBr}), giving rise to a `renormalized                
determinant'. But here we are concerned with something                          
else, namely an infrared divergence: since an instanton configuration has       
nontrivial topology, $V(x)$ behaves as $|x|^{-2}$ as $|x|\to\infty$             
and thus it is an infrared singular perturbation of the Laplacian.              
Just as in nonrelativistic quantum mechanics Levinson's                         
theorem fails for long range potentials, the renormalized                       
determinant of such an operator fails to exist in the infinite volume limit. In 
technical terms \cite{Seiler}, neither ${\cal O}$ nor any finite power of it  
are trace class (actually not even compact) and therefore if one computes the   
(renormalized) determinant for a sphere or a ball of radius                     
$R$, the limit $R\to\infty$ does not exist since the determinant contains       
a term diverging like $\log R$, as found explicitely by Patrascioiu and Rouet   
\cite{PatRouet}\cite{PaRo}.                                                 
                                                                                
\vskip1cm                                                                       
{\it c)  $1/N$ expansion}                                                       
                                                                                
Over the years, many authors have used results obtained in the $1/N$            
expansion as evidence that perturbation theory does provide the correct         
asymptotic expansion (see for instance \cite{FlyvN} and references given      
there). Let us use the usual scaling and write                                  
                                                                                
$$\beta=N\tilde\beta \eqno(2)$$                                                 
                                                                                
\noindent                                                                       
In his well known paper Kupianen \cite{Kupi}                                  
proved that the $1/N$ expansion is                                              
an asymptotic expansion at fixed $\tilde\beta$.                                 
His error estimates are such that                                               
they do not allow to interchange the limits                                     
$\tilde\beta\to\infty$ and $N\to\infty$,                                        
in particular not for long range quantities                                     
such as the correlation length $\xi$.                                           
In fact the numerical data produced by Wolff                                    
\cite{WolffO2,WolffO3,WolffON}                                              
show very clearly that with increasing                                          
$\tilde \beta$ one has to go to                                                 
higher and higher $N$ to achieve a certain                                      
degree of closeness between the correlation                                     
lengths of the $O(N)$ model and the $O(\infty)$                                 
(spherical) model at a given $\tilde\beta$.                                     
                                                                                
So there is no conflict between the                                             
successes of the $1/N$ expansion and the possibility that the                   
$O(N$) model undergoes a phase transition                                       
at some $\tilde\beta_{KT}(N)$. What does follow however from                    
Kupiainen's work (see the introduction of his paper)                            
is that if $\tilde\beta_{KT}(N)<\infty$, then                                   
$\tilde\beta_{KT}(N)$ has to                                                    
grow at least like a power of $\ln N$ for $N\to\infty$.                         
\vskip1cm                                                                       
{\it d) High Temperature Expansions}                                            
                                                                                
Butera, Comi and Marchesini \cite{Butera1,Butera2}                           
computed high temperature series (see also \cite{LuWe})                       
up to order $\beta^{14}$ and Pad\'e approximations to gain                      
insight into the possible singularity structure of                              
the $O(N)$ models in the complex $\beta$-plane.                                 
They conclude that the $O(2)$ model has the nearest                             
singularity on the positive real axis (and identify                             
this singularity with the $KT$ transition), whereas                             
for $N\geq 3$ they find that the closest singularities are                      
off the real $\beta$-axis. This is taken as supporting the absence of           
a phase transition in those models. It should be remarked, however,             
that their results do not provide any evidence against singularities            
on the real axis that are further from the origin than the complex conjugate    
pair they find. Pad\'e approximants are notoriously unable to                   
find such singularities which are `shielded' by closer ones.                    
                                                                                
Bonnier and Hontebeyrie \cite{BonnH} used Pad\'e                              
resummation in a conformally mapped variable that                               
is based on assuming the asymptotic scaling predicted                           
by the perturbative $\beta$-function. They report good                          
agreement with Monte-Carlo data, but unfortunately the                          
data they are using are very old ones of poor quality.                          
The agreement deteriorates markedly if one is using the                         
better data now available \cite{WolffO3}.                                     
                                                                                
There is another point that should be noted: For $N=2,3,4$ the high             
temperature series for the susceptibility has only positive                     
coefficients up to the order to which it has been computed \cite{LuWe}.       
But a power series with positive coefficients has its nearest                   
singularity on the positive real axis; so unless one believes                   
that the coefficients computed so far somehow know already that                 
the higher ones will eventually change sign,                                    
they cannot credibly predict an imaginary part for the                          
closest singularity.                                                            
                                                                                
So it seems fair to say that high temperature expansions,                       
despite their by now remarkable length, have not produced any conclusive        
evidence in favor of the orthodoxy.                                             
\vskip1cm                                                                       
{\it e) `Exact solutions'}                                                      
                                                                                
Zamolodchikov and Zamolodchikov \cite{Zam}                                    
obtained an `exact $S$-matrix' for                                              
the continuum $O(N)$ models under a number of assumptions:                      
First of all they assumed that the model describes                              
an $O(N)$ vector multiplet of massive Bose particles;                           
furthermore they made the usual assumptions about                               
unitarity, analyticity and crossing  symmetry,                                  
and finally the less standard ones of absence of particle creation,             
minimal singularity structure                                                   
and most importantly, factorization.                                            
(Absence of particle creation and factorization are supposed                    
to follow from the existence of infinitely many conservation                    
laws \cite{Polcons,Lucons}                                                   
whose existence, however, in turn depends on some assumptions                   
and cannot be proven without a construction of the                              
continuum limit of the model).                                                  
It is clear that their construction, remarkable as it is,                       
cannot help answer the question of the existence of a                           
massless phase of the lattice models, since it assumes                          
a mass gap from the beginning.                                                  
                                                                                
Using similar assumptions, Karowski and Weisz \cite{KW}                       
derived an `exact current form factor', that is                                 
the matrix element of the current operator between the vacuum                   
and 2-particle states. The same remarks as above apply                          
to this construction.                                                           
                                                                                
Polyakov and Wiegmann \cite{PolWieg} produced an `exact                       
solution' of the $2D$ $O(4)$ model; Wiegmann \cite{Wieg}                      
extended the method to the $O(3)$ model. The first step in                      
this approach is to map the nonlinear $\sigma$-model into a                     
model with 4-fermion interaction which then is to be solved via                 
the Bethe ansatz. The problem is that to establish this                         
equivalence, one has to use an identity for a Gaussian integral                 
over a certain gauge field                                                      
and ignore the fact that actually the gauge fields vary over                    
the compact space $SU(2)=S_3$ and the integration has to be done                
with the Haar measure, not the flat (Lebesgue) measure.                         
Thus the steps required in going from eq.(1) to eq.(2) of \cite{PolWieg}      
cannot be justified on a lattice and are valid only if one                      
imagines taking a naive continuum limit. To quote from a recent                 
paper \cite{FlyvL} it is therefore not clear whether the solution             
`in addition to being exact, is also correct',                                  
and if so, for what model.                                                      
                                                                                
Hasenfratz, Maggiore and Niedermayer \cite{Hasen1} compared                   
the Polyakov-Wiegmann solution that depends on a mass                           
parameter with the perturbation expansion and derived a                         
formula for the mass gap of the $O(3)$ and $O(4)$ models; later                 
Hasenfratz and Niedermayer carried out                                          
a similar calculation for general $O(N)$                                        
\cite{Hasen2} starting from the                                               
Zamolodchikov$^2$ $S$-matrix  and comparing                                     
with perturbation theory. This formula has the property                         
of giving in the limit $N\to\infty$ the correct                                 
asymptotic behavior of the mass gap of the spherical model                      
for $\tilde\beta\to\infty$. So it should come as no surprise                    
that numerical tests (see {\it f)} below)                                       
showed good agreement with the formula for large $N$ but                        
considerable deviations for smaller $N$.                                        
The derivation itself involves some assumptions that                            
may be questioned, such as the description of a very                            
dense gas of particles in terms of a 2-particle $S$-matrix.                     
In addition, as already noted, the $S$-matrix used as an input can only be      
considered a clever guess, considering the many assumptions that                
are needed to obtain it, and its connection with                                
any $O(N)$ lattice model is not in the least transparent.                       
So whatever its merits, this work also does not contribute                      
to answer the question of the mass gap, since its existence                     
has to be assumed from the start.                                               
\vskip1cm                                                                       
{\it f) Numerical results}                                                      
                                                                                
Numerous papers have appeared reporting Monte Carlo                             
investigations that are claimed to                                              
support the absence of a phase transition in the standard nearest               
neighbor action (s.n.n.a.) non-Abelian ferromagnets \cite{Shenker,Fukugita,  
Heller,Petronzio,Bernreuther}.                                               
We believe that to a large extent these                                         
claims were motivated by the authors'                                           
expectations and that in fact an                                                
objective analysis of the numerical                                             
situation suggests rather the contrary.                                         
Namely, there is universal agreement                                            
that in almost all the standard action                                          
models there is a `cross-over' region,                                          
where the magnetic susceptibility and                                           
the correlation length increase faster                                          
than the  asymptotic freedom predictions.                                       
This `cross-over' region is supposed to reflect                                 
the existence of a line of first                                                
order transitions, terminating at a critical                                    
point in the `mixed action models' parametrized by                              
$(\beta_1,\beta_2)$, where                                                      
                                                                                
 $$H(\beta_1,\beta_2)=-\sum_{\langle xy\rangle}                                 
 \{\beta_1S(x)\cdot S(y)+\beta_2(S(x)\cdot S(y))^2\} \eqno(3)$$                 
                                                                                
We think that this scenario is highly implausible; indeed although not          
rigorously proven, one would expect that in a ferromagnet susceptibility and    
correlation length are nondecreasing functions of $\beta_1$                     
and $\beta_2$. Hence if                                                         
they diverge at the point $(\beta_{1,KT},\beta_{2,KT})$,                        
they must continue to do so in the                                              
whole region $\beta_1\geq\beta_{1,KT}$ and $\beta_2\geq\beta_{2,KT}$.           
But there are no phase boundaries separating                                    
this region from the line $\beta_2=0$. Hence the critical region must touch     
the line $\beta_2=0$. Therefore we believe                                      
that there is no good explanation for the repeated                              
occurrence of the so called `cross-over' region                                 
observed in both $2D$ $O(N)$ models and $4D$                                    
gauge theories. On the other hand, it could very well be                        
that this region is not a cross-over, but rather the                            
neighborhood of a critical point, which the orthodoxy claims                    
should not exist.                                                               
                                                                                
We would like to clear another erroneous belief expressed in                    
many numerical studies. Several authors                                         
\cite{Heller,Petronzio,Hasen0}  have advocated going past                   
the `cross-over' region by employing the                                        
Monte Carlo renormalization group (MCRG)                                        
or some finite size scaling curves.                                             
Common to all such approaches is the belief                                     
that one can take measurements                                                  
on small lattices and learn about the infinite                                  
volume behavior. As we explained                                                
in (a) above, perturbation theory is suspect                                    
precisely because an ordered state does not                                     
exist on an infinite                                                            
lattice. On the other hand, given the                                           
size of the lattice, one can always choose                                      
a $\beta$ sufficiently large so that                                            
all these lattice Green's functions                                             
agree with their perturbative values to                                         
any degree of accuracy (since perturbation theory                               
is clearly asymptotic in a finite volume).                                      
We have checked that already for $\beta$                                        
in the crossover region the finite                                              
volume susceptibility computed via Monte                                        
Carlo agrees with the perturbative                                              
formula two-loop formula given by Hasenfratz \cite{Hasenf}                    
within a few percent, provided the size of the lattice is                       
less than the infinite volume correlation                                       
length, and the agreement is rapidly                                            
improving with increasing $\beta$.                                              
So there is really no insight                                                   
to be gained by running Monte Carlo simulations in this regime.                 
The real dilemma is, do the limits $L\to\infty$ and                             
$\beta\to\infty$ commute? Techniques such as                                    
MCRG not only cannot answer this                                                
question but, by insisting on working on small                                  
lattices, are bound to reproduce                                                
all perturbative predictions.                                                   
                                                                                
L{\"u}scher and Wolff \cite{Luwo} studied numerically                         
the current form factor and the 2-particle                                      
scattering phases of the $O(3)$ model and claimed agreement                     
with the results of Zamolodchikov$^2$ \cite{Zam} and                          
Karowski and Weisz \cite{KW}, respectively.                                   
They worked at values of $\beta$                                                
where the model is clearly in its massive phase and                             
the correlation length is between $6.9$ and $13.6$ lattice units.               
The form factor is fixed by a normalization                                     
condition at zero momentum; for increasing momenta                              
they find increasing deviations from the predicted values. According            
to them the differences can be understood as lattice corrections                
(a derivation of those corrections is not given).                               
They also find that the 2-particle scattering phases                            
roughly agree with the predictions of \cite{Zam}                              
for energies that are small compared to the mass.                               
Discrepancies beyond the numerical accuracy                                     
are again blamed on `lattice                                                    
artefacts' (because these discrepancies were increasing                         
with the energy and were smaller at the smaller value of the                    
two values of the mass investigated).                                           
It should be remarked that the numerical determination                          
of the scattering phases depends on quite a lot                                 
of additional theoretical input and also on an                                  
extraneous parameter (called $t_o$).                                            
But it is most important to realize that by                                     
its very nature such a test cannot answer the question of                       
the existence of a mass gap,  and due to the limitations                        
of numerics it could also not verify the interesting                            
feature of the Zamolodchikov$^2$ $S$-matrix that the                            
high energy limit of the scattering phases is zero,                             
which has been interpreted as a manifestation of                                
asymptotic freedom.                                                             
                                                                                
Wolff \cite{WolffON} performed a numerical study                              
to test the validity of the mass formula of                                     
Hasenfratz and Niedermayer \cite{Hasen2} mentioned above                      
for $N=3,4,8$, of course                                                        
in a regime where there is undoubtedly a mass gap.                              
It turns out the the agreement is not very good  for                            
$N=3$ (between 25\% and 33\%                                                    
discrepancy in the region beween $\beta=1.4$ and $1.9$),
but seems to improve with increasing $N$.                                       
In our opinion this study has no bearing on the                                 
question of the existence of a massless phase, since the data were              
taken in the same range of $\tilde\beta=\beta/N$                                
for the different values of $N$.                                                
Of course for $N\to\infty$ at fixed and sufficiently large $\tilde\beta$        
the agreement has to improve because, as remarked above,                        
the Hasenfratz-Niedermayer formula is correct for the                           
spherical model in the limit $\tilde\beta\to\infty$,                            
and at fixed $\tilde\beta$ the mass gap of the                                  
$O(N)$ model seems to converge to the spherical model                           
mass (Kupiainen \cite{Kupi} proved that the mass gap                          
is bounded below by a quantity converging to the  spherical                     
model mass). But we know anyway from Kupiainen's work                           
\cite{Kupi} that $\tilde\beta_{KT}$                                           
has to increase at least logarithmically with $N$.                              
So only a study of the mass gap in such a regime                                
could possibly give any information about the                                   
existence or nonexistence of a massless phase;                                  
this is beyond the present numerical possibilities                              
even with the new  cluster algorithms.                                          
\vskip1cm                                                                       
{\it g) The role of topology}                                                   
                                                                                
The original Kosterlitz-Thouless scenario was that in the $O(2)$ model          
a phase transition must occur, reflecting the loss of topological order as      
the temperature is increased. Their conclusion was derived from the             
energy-entropy arguments mentioned earlier:                                     
on a lattice of size $L$ a vortex has an energy of                              
order $\log L$. Its entropy, measuring essentially the location of the center   
is also $O(\log L)$. Hence if $\beta$ is too large, vortices are bound, while at
high temperature they unbind, triggering the phase transition. These            
considerations suggest also a basic difference between $O(2)$ and $O(3)$;       
in the latter                                                                   
smooth configurations -- instantons -- have energies $O(L^0)$, hence they       
act like point defects and disorder the system                                  
at arbitrarily low temperatures.                                                
                                                                                
To understand better the role of topology, we found it useful                   
to consider a modification of the                                               
$O(2)$ model dubbed `cut' model in \cite{PS1}, 'constrained model'            
in \cite{PS2}:                                                                
the Gibbs factor retains its                                                    
form only for $|S(x)-S(y)|\le \epsilon$, while it                               
is replaced by $0$ otherwise. Thus we are forbidding                            
large angular deviations between neighboring                                    
spins. Since this modification is ferromagnetic,                                
we would expect                                                                 
$\beta_{KT}$ to decrease as $\epsilon$ is decreased from its                    
s.n.n.a. value of $2$. This is indeed                                           
what we observed numerically. The surprising result though was that             
we found that for about $\epsilon=1.57$, $\beta_{KT}=0$                         
on a square lattice; for this value of $\epsilon$ vortices are still            
allowed and at $\beta=0$ they cost no energy!                                   
This finding suggests                                                           
that the original Kosterlitz-Thouless argument                                  
was too naive in its estimate                                                   
of the entropy: at $\beta=0$, the Gibbs factor                                  
is either $1$ (for $|S(x)-S(y)|\le \epsilon$) or 0                              
(otherwise). So on a square lattice,                                            
the Kosterlitz-Thouless argument would                                          
have suggested that $\beta_{KT}=0$ only for                                     
$\epsilon\leq\sqrt 2$, since that is the value for which                        
vortices cease to exist.                                                        
                                                                                
The low temperature phase of the $O(2)$ model must thus                         
be characterized by the fact that vortices                                      
are sufficiently rare and spin waves dominate.                                  
It was suggested already by one of us in \cite{Pat87}                         
that in both Abelian and Non-abelian models at large                            
$\beta$ spin waves may dominate and defects may be suppressed.                  
In \cite{Pattex} one of us is giving more detailed arguments                  
in favor of this scenario. According to this scenario the situation             
would be as follows: just as in $D\ge 3$                                        
the very good `local order' present at large $\beta$ manifests itself           
thermodynamically as l.r.o., in $2D$ its manifestation may be algebraic order.  
As the temperature is raised, defects --                                        
bonds where $|S(x)-S(y)|$ is large -- become more                               
abundant and at some point condensate, putting the system in a phase            
characterized by exponential decay of correlation functions. Artificially       
suppressing defects should have an effect in all dimensions and we predict      
that a suitably modified $O(N)$ model will exhibit l.r.o. at all $\beta>0$      
in any $D\ge 3$.                                                                
\vskip1cm                                                                       
                                                                                
{\it h) $QCD_4$}                                                                
                                                                                
One may wonder what this discussion implies for $QCD$. It has generally been    
accepted that there is a close analogy between $2D$ spin models and             
$4D$ gauge models. Most of our discussion applies there equally well,           
with the proviso that the evidence for the orthodoxy is weaker:                 
the $1/N$ expansion does not have a firm base like Kupiainen's \cite{Kupi}    
results, there is no exactly soluble $N=\infty$ limit,                          
there are no `exact solutions' for the $S$-matrix, the                          
role of topology for disordering the system is even less clear and              
finally the numerical results are much more limited by the increased            
requirement of computing power.                                                 
                                                                                
The existing numerical data certainly cannot rule out a deconfining             
phase transition and the famous 'dip' in the $\beta$-function may               
in fact be an indication of its presence, since it shows that                   
quantities like the string tension tend to zero faster than                     
asymptotic scaling would predict, just as in the s.n.n.a. $O(N)$                
models in the so-called cross-over region. The usual explanation, that          
the 'dip' is a cross-over regime induced by the presence of a critical          
point in the $\beta_{fund}-\beta_{adj}$ plane \cite{BhanotD}, could           
again be ruled out by correlation inequalities, hence seems implausible.        
                                                                                
Finally, let us recall the following bizarre situation: it is known             
that all gauge theories undergo a deconfining transition at                     
finite temperature \cite{BorgsS}.                                             
It is also agreed that $U(1)$ and $SU(N)$, $N\geq 4$ gauge                      
theories undergo a transition at zero temperature. For the Abelian case         
$U(1)$ that transition is known rigorously \cite{FS,GU}                      
to be deconfining. For the non-Abelian cases $SU(N)$, $N\geq 4$, the claim is   
that the transition is first order and not deconfining \cite{Creutz1,Creutz2 
,Moriarty,BohrM,CreutzM}. It has been recognized \cite{DasK1,DasK2}
that it is difficult to separate this first order `bulk' transition from        
the finite temperature deconfining transition, because                          
the strange thing is that the extrapolation of the curve describing             
the deconfining transition in the $\beta-T$ plane ($T$ is the temperature)      
seems to cross the axis $T=0$ just where the supposed first order transition    
occurs.                                                                         
In the orthodox outlook, these                                                  
two transitions are supposed to have nothing to do with each other              
(in fact we know of no explanation for the first order transition at            
zero temperature, which for instance is not seen in $SU(2)$ and $SU(3)$). Then  
one could ask: if in $SU(N)$, $N\geq 4$, these two transitions have nothing to  
do with each other, should it not be possible to change the action              
and pull the two transitions apart, perhaps even eliminating the first          
order transition altogether?                                                    
Could it not be that the difficulty of separating the two                       
transitions simply means that the first order transition at zero temperature is 
nothing but the deconfining transition? This possibility was discussed in       
\cite{PSSLinke}.                                                              
\vskip1cm                                                                       
\centerline{{\bf 4. Rigorous results for a modified O(N) model}}                
\vskip5mm
To illustrate that the existence of instantons  and perturbative                
asymptotic freedom do not rule out nonanalytic behavior we                      
consider the following model which is                                           
inspired by some considerations for lattice gauge theories                      
due to Mack and Petkova \cite{MP}:                                            
It is defined by the following modification of the s.n.n.a.\ measure:           
for every plaquette $p$, we require                                             
                                                                                
$$\prod_{\langle xy \rangle\in\partial p}                                       
sgn(S(x)\cdot S(y))=1   \eqno   (4)$$                                           
                                                                                
\noindent                                                                       
where the product is over the bonds around the plaquette $p$.                   
                                                                                
{\it Remark}: This modification should be                                       
unimportant at low temperatures. Indeed,                                        
using a result of Bricmont and Fontaine \cite{BrFo},                          
the probability to violate the                                                  
constraint at a given plaquette in the s.n.n.a.                                 
$O(N)$ model is bounded by $\exp(-c\beta)$ for some $c>0$.                      
Moreover the modification does not affect the                                   
existence of asymptotic freedom in perturbation theory nor the presence of      
instantons (defects costing an energy of order 1) in $O(3)$.                    
Finally we remark that the `cut' or `constrained' models                        
discussed in \cite{PS1,PS2} (see also point {\it h)} of the previous         
section), in which the angular deviation between neighboring                    
spins is limited, automatically satisfy the constraint (4).                     
We note that the constraint should make the system more ferromagnetic           
so that one would expect                                                        
                                                                                
$$\langle S(0)\cdot S(x)\rangle_m > \langle S(0)\cdot S(x)\rangle               
_{s.n.n.a.}\eqno(5)$$                                                           
                                                                                
\noindent                                                                       
where $\langle . \rangle_m$ denotes the expectation in the                      
modified model.                                                                 
For this  model  one can introduce the disorder parameter                       
                                                                                
$$\langle D(x)\rangle_m=\lim_{L\to\infty}                                       
\langle\prod_{k=0}^{L-1}\exp\bigl(-2\beta sgn(S(x_k)                            
\cdot S(x_{k+1}))\bigr)\rangle_m                                                
\eqno(6)$$                                                                      
                                                                                
\noindent                                                                       
where the product is over a path (ordered set of bonds) from                    
$x$ to $x_L$. Because of the constraint (4) $\langle D(x)\rangle_m$ is
path independent.                                                            
The following theorem can be proven along the lines                             
of Mack and Petkova \cite{MP}:                                                
                                                                                
{\bf Theorem}:                                                                  
                                                                                
a) There exists a $\beta_1\in(0,\infty)$,                                       
such that for $\beta<\beta_1$                                                   
$$\langle D(x)\rangle_m >0$$                                                    
                                                                                
b)There exists a $\beta_2\in(\beta_1,\infty)$                                   
such that for $\beta>\beta_2$                                                   
                                                                                
$$\langle D(x)\rangle_m=0$$                                                     
                                                                                
\noindent                                                                       
{\it Remark}: The only reason why one could doubt that                          
this theorem implies the existence of a phase                                   
transition in these $O(N)$ models is                                            
the non-local nature of the disorder                                            
variable $D(x)$. But the general experience gathered                            
in the study of numerous models is that a disorder                              
operator of the kind used here signals correctly the occurrence                 
of a phase transition and can even be used to determine                         
exactly its location.                                                           
                                                                                
{\it Proof:} We do not give a detailed proof, because                           
the adaptation of the proof given by Mack and Petkova                           
for gauge theories is straightforward. We will only present                     
the strategy.                                                                   
                                                                                
Because of the constraint (4) one can define Ising spins                        
$\{\sigma_x\}$ attached to the sites as follows:                                
                                                                                
$$\sigma_x=\prod_{k=0}^{L-1} sgn\bigl(S(x_k)\cdot S(x_{k+1})\bigr)              
 \eqno(7)$$                                                                     
                                                                                
\noindent                                                                       
where as before the poduct is over a path (of length $L$) from                  
$0$ to $x$ and by (4) there is no dependence on the path.                       
The system can now be rewritten as a coupled system of the                      
$O(N)$ spins $S$ and the Ising spins $\sigma$, described by the                 
Hamiltonian                                                                     
                                                                                
$$H=\sum_{\langle xy\rangle}\sigma_x\sigma_y S(x)\cdot S(y)                     
\eqno(8)$$                                                                      
                                                                                
\noindent                                                                       
with the constraint that for all bonds $\langle xy\rangle$                      
\ $S(x)\cdot S(y)\ge 0$.                                                        
As in \cite{MP} one carries out a duality transformation                      
of the Ising variables, leaving the $S$ variables alone.                        
One obtains again a certain Ising model with fluctuating                        
coupling constants given by                                                     
                                                                                
$$\tilde\beta_{xy}=-{1\over 2}\ln\tanh\Bigl(\beta (S(x)\cdot S(y) \Bigr)        
\eqno(9)$$                                                                      
                                                                                
\noindent                                                                       
Then as in \cite{Richard}  the GKS correlation inequalities imply that        
                                                                                
$$\langle D(x)\rangle_m\ge |\langle D(x)\rangle_{Is}|                           
\eqno(10)$$                                                                     
                                                                                
\noindent                                                                       
where the right hand side denotes the expectation value of the standard         
disorder operator in the Ising model  at the same value of $\beta$.             
(10) simply expresses the fact that the coupling of the Ising variables         
in the model described by (8) is less ferromagnetic than in the standard        
Ising model at the same $\beta$. So we have learned that for                    
$\beta \ge\beta_{c,Is}={1\over 2}\ln(\sqrt{2}+1)                                
\quad \langle D(x)\rangle_m=0$. In other words                                  
the first part of the theorem is proven for                                     
                                                                                
$$\beta_1\ge {1\over 2}\ln(\sqrt{2}+1) \eqno(11)$$                              
                                                                                
As is well known, the disorder parameter will become                            
the two-point function of the Ising spins after the                             
duality transformation.                                                         
The crucial point is that $\tilde\beta_{xy}$ will typically be                  
very small, in other words, the dual model will be in its                       
high temperature phase, and its 2-point function will                           
decay exponentially. The actual proof proceeds by                               
cluster expanding the conditional expectation value                             
of the dual Ising two point function for fixed                                  
couplings $\{\tilde\beta_{xy}\}$. One takes the remaining                       
(annealed) expectation value over these couplings                               
termwise in the cluster expansion. These expectations                           
can be bounded above using  the so-called chessboard                            
bounds (after undoing the duality transformation),                              
since the measure still possesses  reflection                                   
positivity. In this way one obtaines convergence of                             
the cluster expansion and exponential  decay of                                 
$\langle D(x)\rangle_m $. For details of this proof                             
we refer the reader to the paper of Mack and Petkova                            
\cite{MP}.                                                                    
                                                                                
We would like to remark that something more is true                             
for this model:                                                                 
the Ising spins (7) of this model show a transition                             
from a disordered high temperature phase to a phase with                        
l.r.o.. The first part of this remark is a direct consequence of                
the GKS inequalities. To prove the second part one has to adapt                 
Georgii's result \cite{GeorgiiCMP} on                                         
the percolation of the low energy bonds                                         
to this model; this result implies in particular that                           
the bonds $\langle xy\rangle$ with $sgn\bigl(S(x)\cdot S(y)\bigr)>0$            
percolate. Using the definition (7) one sees that                               
this implies the existence of a percolating cluster                             
of + Ising spins, implying long range order.                                    
                                                                                
\vskip1cm                                                                       
{\it Acknowledgment:} We are grateful to the Max-Planck-Institut f\"ur          
Physik and the Center for the Study of Complex Systems of the                   
University of Arizona for their continued support of the research               
reported here and we thank F.Niedermayer and P.Weisz for reading the            
manuscript.                                                                     
  
\end{document}